\documentclass[fleqn,12pt]{wlscirep}
\usepackage[utf8]{inputenc}
\usepackage[T1]{fontenc}
\title{Magnon gap excitations in van der Waals antiferromagnet MnPSe$_3$}

\author[1,*]{Dipankar Jana}
\author[1]{D.~Vaclavkova}
\author[1]{I.~Mohelsky}
\author[1,2]{P.~Kapuscinski}
\author[1]{C.~W.~Cho}
\author[1]{I.~Breslavetz}
\author[3,4]{M.~Bia\l{}ek}
\author[4]{J.-Ph.~Ansermet}
\author[1]{B.~A.~Piot}
\author[1,5]{M.~Orlita}
\author[1]{C.~Faugeras}
\author[1,2,3,$\dag$]{M.~Potemski}

\affil[1]{Laboratoire National des Champs Magn\'etiques Intenses, LNCMI-EMFL, CNRS UPR3228,Univ. Grenoble Alpes, Univ. Toulouse, Univ. Toulouse 3, INSA-T, Grenoble and Toulouse, France}
\affil[2]{Institute of Experimental Physics, Faculty of Physics, University of Warsaw, ul. Pasteura 5, PL-02-093 Warsaw, Poland}
\affil[3]{CENTERA Labs, Institute of High Pressure Physics, PAS, 01-142 Warsaw, Poland}
\affil[4]{Institute of Physics, Ecole Polytechnique Fédérale de Lausanne (EPFL), 1015 Lausanne, Switzerland}
\affil[5]{Institute of Physics, Charles University,  Prague, 121 16, Czech Republic}

\affil[*]{dipankar.jana@lncmi.cnrs.fr}
\affil[$\dag$]{marek.potemski@lncmi.cnrs.fr}

\begin{abstract}
Magneto-spectroscopy methods have been employed to study the zero-wavevector magnon excitations in MnPSe$_3$.  Experiments carried out as a function of temperature and the applied magnetic field show that two low-energy magnon branches of MnPSe$_3$ in its antiferromagnetic phase are gapped. The observation of two low-energy magnon gaps (at 14 and 0.7 cm$^{-1}$) implies that MnPSe$_3$ is a biaxial antiferromagnet.  A relatively strong out-of-plane anisotropy imposes the spin alignment to be in-plane whereas the spin directionality within the plane is governed by a factor of 2.5 $\times$ 10$^{-3}$ weaker in-plane anisotropy.
\end{abstract}
\begin{document}

\flushbottom
\maketitle
%
%
\thispagestyle{empty}

\noindent

\section*{Introduction}

Two-dimensional (2D) materials beyond graphene constitute an exceptional platform for the exploration of the unique electronic, optical, and magnetic properties of low-dimensional systems \cite{bhimanapati2015, manzeli20172, mak2016, geng2018, molle2017}. Among currently studied materials are layered, van der Waals magnets, which include a family of metal phosphorus trichalcogenides (MPX$_3$, with M = Mn, Fe, Co, Ni and X = S, Se)\cite{Xing2019, Mai2021, Liu2021, Cui2023, Wildes2023, Kang2020}. MPX$_3$ compounds are semiconducting antiferromagnets with diverse spin ordering configurations of the transition metal ions \cite{Brec1986}. Since MPX$_3$ layers are held together by weak van der Waals forces, typical distances between the metal ions within the layers are shorter than those between ions in different layers, and the magnetic order in MPX$_3$ compounds is governed mainly by spin-spin coupling within the layers. Such quasi-2D or even strictly 2D (in monolayers) magnetic ordering persists in MPX$_3$ materials since they are not necessarily perfect Heisenberg antiferromagnets, but, in addition to the exchange terms, are also characterized by non-zero magnetic anisotropies. A non-negligible spin-orbit interaction within the M-ions is a primary cause of the so-called single-ion magnetic anisotropy. At first sight, the spin-orbit interaction is expected to be rather weak for Mn$^{2+}$ ions in MnPX$_3$, though the nature seems to be against the Mermin-Vagner theorem \cite{Mermin1966} (no magnetic order in the 2D limit for Heisenberg antiferromagnets) and MnPX$_3$ compounds do display antiferromagnetic ordering \cite{joy1992, Jeevanandam1999}. Interestingly enough, the change of the chalcogen atom from sulfur (S) to selenium (Se) imposes the change in the type of magnetic ordering \cite{Basnet2022, Han2023}. A single-ion anisotropy may still be relevant for ordering spins in MnPSe$_3$ \cite{Jeevanandam1999, Basnet2022}, whereas the competing anisotropy due to magnetic dipolar interaction is likely imposing the character of spin-ordering in MnPS$_3$ \cite{joy1992}. From another perspective, the single ion anisotropy in MnPSe$_3$ is considered to be small while interlayer exchange integrals are significantly large \cite{Calder2021}. Despite these ambiguities, it is known that spins are aligned in the direction nearly perpendicular to the layers' plane in the case of broadly studied MnPS$_3$ \cite{Goossens2000}. Instead, the less studied MnPSe$_3$ is an easy-plane antiferromagnet, as deduced from neutron scattering \cite{wiedenmann1981, Calder2021} and magnetization \cite{Jeevanandam1999, Basnet2022} measurements as well as theoretical calculations \cite{Mai2021}. The open questions addressed in the present work are in regards to the character of low energy magnon excitations in MnPSe$_3$. One of the calculated magnon branches~\cite{Mai2021} has been concluded to be gapless (being dispersed linearly with the $k$-vector) and the $k=0$ gap of the second magnon branch is estimated to be $\sim$14~cm$^{-1}$. Neutron scattering data\cite{Calder2021} confirm the upper ~14~cm$^{-1}$ gap but lack the resolution to trace the low energy magnon branch in the $k=0$ limit. On the other hand, the magnetization measurements \cite{Basnet2022} indicate a spin-flop field at $\sim$0.7~T, pointing out a non-zero onset (at $k=0$) of the lower magnon branch.

In the present work, we demonstrate that magnon gap excitations in MnPSe$_3$ antiferromagnet can be effectively probed using magneto-spectroscopy methods (Raman scattering, far-infrared, and microwave absorption techniques). With these experiments, we identify MnPSe$_3$ as a \textit{de facto} biaxial antiferromagnet, by revealing the energies of two fundamental magnon gaps, both of which are non-zero (14 and 0.7~cm$^{-1}$). This implies that the antiferromagnetic order in MnPSe$_3$ is governed by two anisotropy terms: one previously identified and directed across the layers' plane ($D_{\perp} \approx 0.3$~cm$^{-1}$)\cite{Basnet2022}, which imposes the in-plane spin alignment, and another one, weak but non-negligible $D_{\parallel}$ term, estimated here to be $\approx 7.5 \times 10^{-4}$~cm$^{-1}$, which imposes the spins' alignment along a certain, yet unidentified, direction within the plane.

\section*{Results}

\begin{figure}[ht]
\centering
\includegraphics[width=8.4cm]{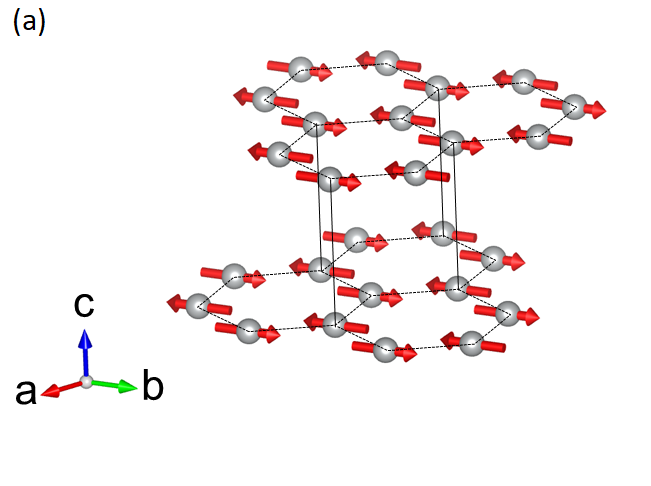}
\includegraphics[width=8.4cm]{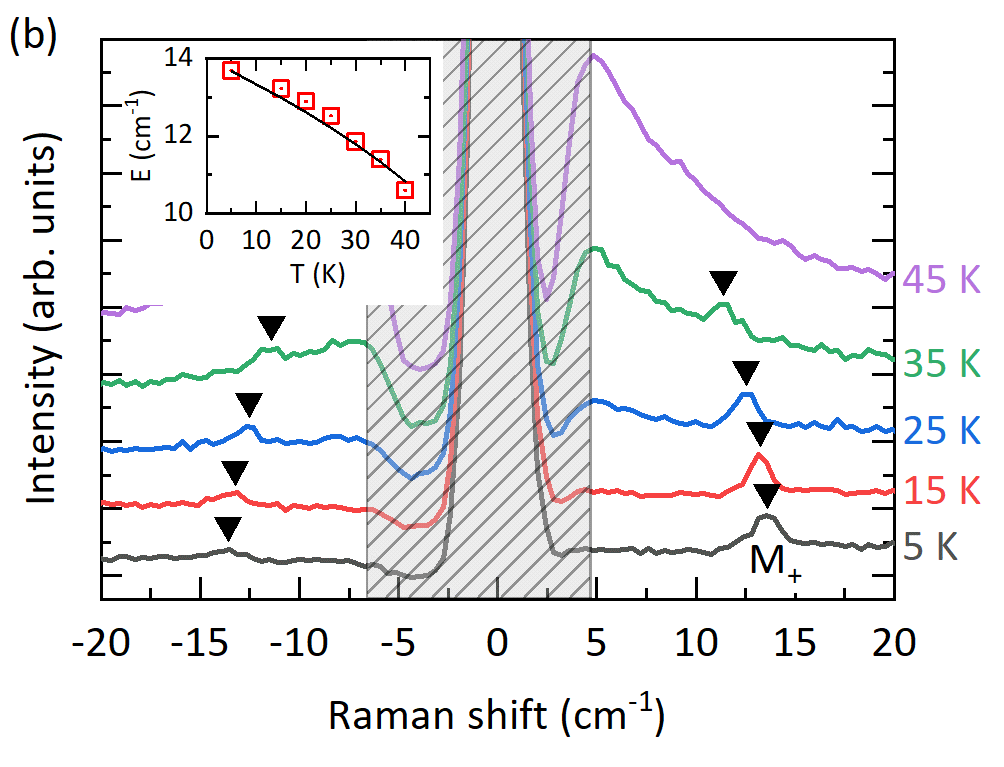}
\caption{ (a) The magnetic structure of MnPSe$_3$ in the antiferromagnetic phase. The grey sphere and red arrow represent Mn$^{2+}$ ion and spin direction respectively. The figure is created using the VESTA software package. \cite{Momma2011}. (b) Raman spectra of MnPSe$_3$ measured at some selected temperatures. M$_+$ resonance corresponds to the upper magnon gap excitation. Spectra are shifted vertically for clarity. Black triangles mark the peak energy $w_{M+}$ of M$_+$ resonance. The shaded region indicates the spectral range blocked by filtering of the laser stray light. As illustrated in the inset, the $w_{M+}$ dependence on temperature follows a conventional formula for the antiferromagnets (see the text) as shown with the solid line.}
\label{fig:Fig1}
\end{figure}

\begin{figure}[ht]
\centering
\includegraphics[width=8.4cm]{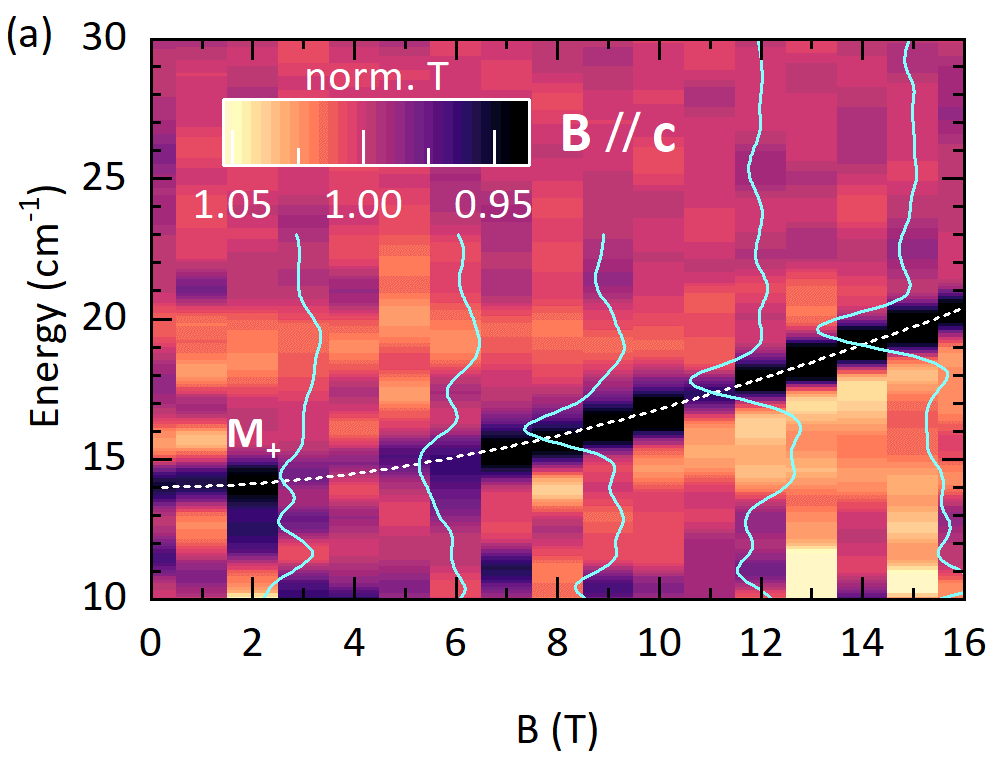}
\includegraphics[width=8.4cm]{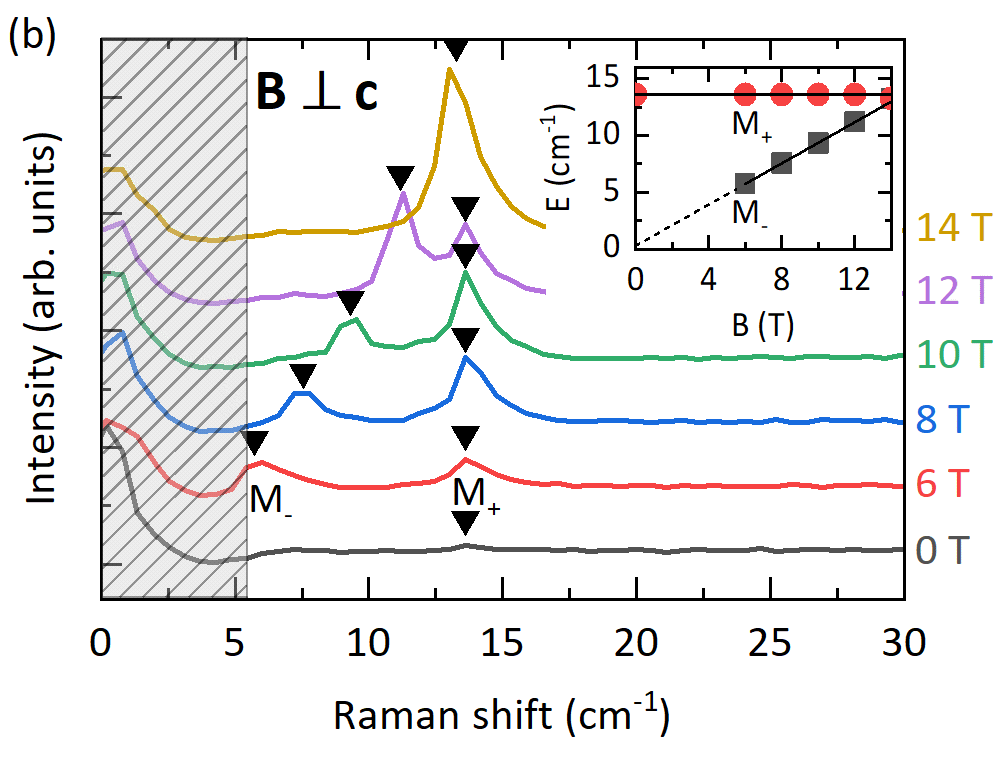}
\caption{(a) False color map of normalized far-infrared transmission of MnPSe$_3$, measured at 4.2~K, as a function of the magnetic field applied along the crystal c-axis. A few representative spectra, measured at a magnetic field strength of 3~T, 6~T, 9~T, 12~T, and 15~T are also plotted. The dashed line corresponds to a fit of $w_{M+} (B)$ dependence according to Eq. \ref{eq:1}. (b)  Raman scattering spectra of MnPSe$_3$ measured at selected in-plane magnetic fields. M$_+$ and M$_-$ resonances are identified as, correspondingly the upper and lower magnon gap excitations. Spectra are shifted vertically for clarity. Black triangles mark the peak energies which are plotted as a function of magnetic field in the inset. The solid line corresponds to a linear fit to the field dependence of M$_-$ and M$_+$ modes while the dotted line corresponds to extrapolation of the fit to 0~T. The shaded region indicates the spectral range blocked by filtering of the laser stray light.}
\label{fig:Fig2}
\end{figure}

The magnetic structure of MnPSe$_3$ in the antiferromagnetic phase is shown in Fig.~\ref{fig:Fig1}a. It is assumed \cite{Jeevanandam1999} that the in-plane alignment of the colinear Mn$^{2+}$ magnetic moments is imposed by the (single ion) anisotropy term  ($D_{\perp} \approx 0.3$~cm$^{-1}$)\cite{Basnet2022} along the c-axis. The axis of spin alignment within the a-b plane is chosen arbitrarily. So far, no in-plane anisotropy term was theoretically identified in MnPSe$_3$.  The net antiferromagnetic order in the planes is governed by the nearest neighbor exchange interactions. The spin alignment across the layers is reported to be antiferromagnetic  \cite{Calder2021}.
Raman scattering is one of the experimental techniques used to probe the low energy excitations in solids and in particular the $k=0$ spin excitations (magnon gap excitations) in layered antiferromagnets\cite{balkanski1987}. Such a method revealed the coupled phonon-two magnon excitations in MnPSe$_3$ antiferromagnet\cite{Mai2021} which appear in the spectral range relatively far from the laser line. Instead, in our experiments, we focus on the Raman scattering response of MnPSe$_3$ in an energy range close to the laser line. The temperature evolution of Raman scattering spectra of MnPSe$_3$ measured in the range of $\pm$~20~cm$^{-1}$ is shown in Fig.~\ref{fig:Fig1}b. The Raman scattering feature at an energy of $w_{M+}\approx14$~cm$^{-1}$ in the limit of low temperatures, denoted as M$_+$, characteristically softens with rising temperature, is identified as an upper magnon gap excitation of MnPSe$_3$. Notably, the $w_{M+}$ energy coincides well with the previously reported value \cite{Mai2021}. The optional approach to trace the temperature evolution of the M$_+$ mode, with the technique of THz monochromatic transmission spectroscopy in the spectral range $270-390$ GHz ($9-13$ cm$^{-1}$), is presented in section S1 of the SM\cite{SuppInfo}.

\begin{figure}[ht]
\centering
\includegraphics[width=8.4cm]{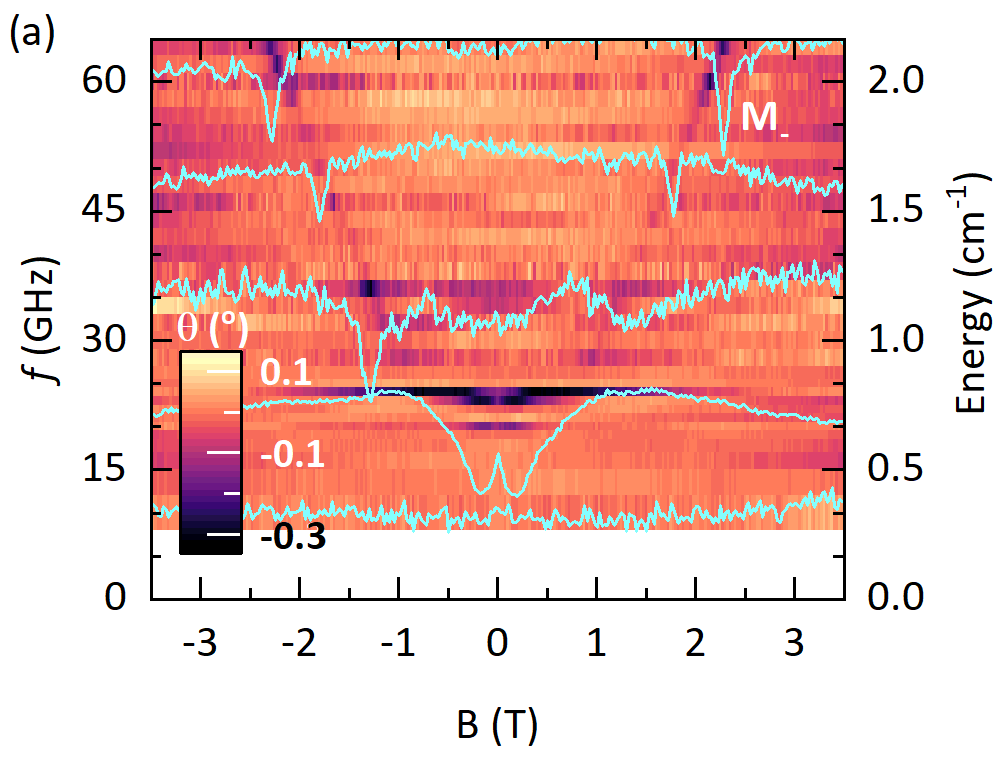}
\includegraphics[width=8.4cm]{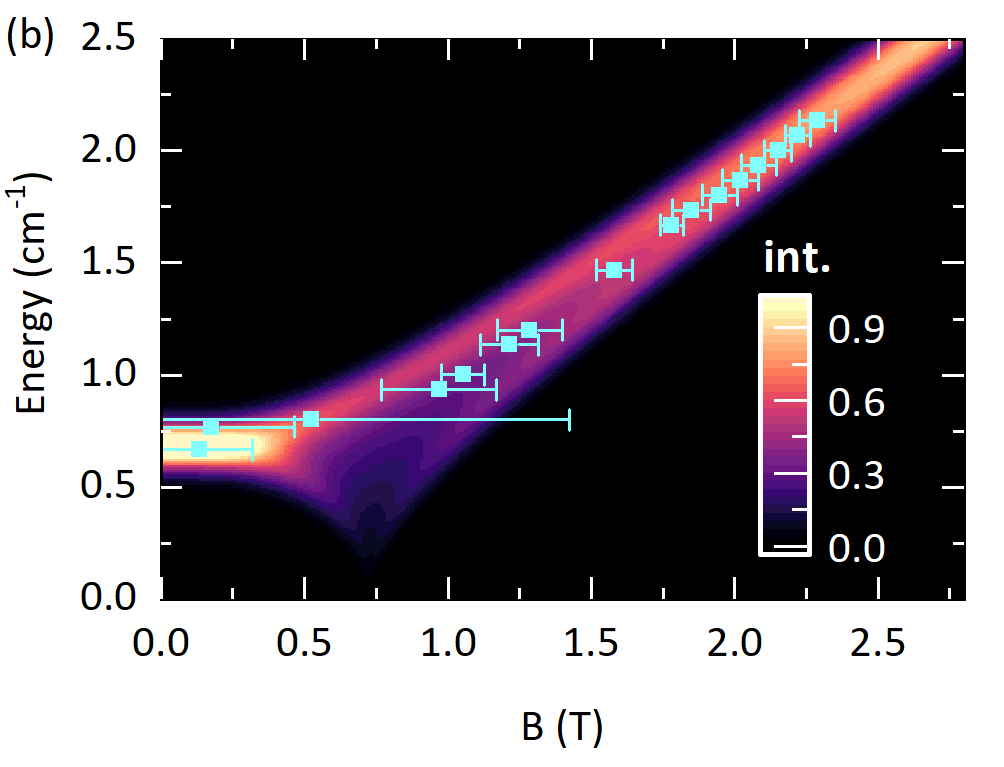}
\caption{ (a) False color map of GHz absorption phase of MnPSe$_3$, measured at 5~K, as a function of the in-plane magnetic field. A few spectra at selected frequencies are also displayed in the same plot showing a change in the absorption phase at the resonance energy. (b) False color map of the simulated (normalized) intensity of M$_-$ mode resonance absorption considering the uniform distribution (from 0$^\circ$ to 90$^\circ$ with respect to the in-plane magnetic field) of spin domains in the plane. Peak energies obtained from phase spectra (Fig.~\ref{fig:Fig3}a) are shown by cyan squares whereas the half-widths of the observed resonances are indicated with horizontal bars.}
\label{fig:Fig3}
\end{figure}

Next, we focus on experiments carried out at low temperature, but as a function of the magnetic field applied either in the direction perpendicular or parallel to the layers' planes (out-of-plane and in-plane configurations, correspondingly). The evolution of the $M_+$ mode with the magnetic field applied in the out-of-plane configuration could be conveniently traced with far-infrared transmission measurements. The results of these experiments are shown in Fig~\ref{fig:Fig2}a. We observe that the $w_{M+}$ increases upon applying the magnetic field. One may anticipate that $w_{M+}$ initially displays a quadratic $B$ dependence, which turns into a linear one at high magnetic fields. Our approach to trace the M$_+$ by Raman scattering in the out-of-plane configuration was not successful due to the specific arrangement of our experimental set-up, preventing an efficient rejection of the stray laser light. Nevertheless, we performed such experiments by setting an angle of $20^{\circ}$ between the direction of the magnetic field and the normal to the surface. The results of these experiments, as shown in section S2 of the SM,\cite{SuppInfo} are consistent with those presented in Fig~\ref{fig:Fig2}a.
The performance of our magneto-Raman scattering set-up is superior in the configuration of the in-plane magnetic field.  A series of low-temperature (5~K) Raman scattering spectra measured as a function of the in-plane magnetic field (arbitrary direction within the layers’ plane) is shown in Fig.~\ref{fig:Fig2}b. The energy of the M$_+$ mode is found to be unaffected by the in-plane magnetic field. On the other hand, another low energy  M$_-$ mode emerges in the spectra from the region covered by the stray laser light, when the strength of the applied field exceeds $B=6~T$. In this range of the magnetic field, the energy of the M$_-$ mode is observed to increase linearly with $B$. Interestingly, at fields around $B=14~T$, the M$_-$ mode crosses the M$_+$ mode and $w_{M-}$ > $w_{M+}$ in the range of high magnetic fields (see section S3 of the SM\cite{SuppInfo}). Of particular interest for defining the character of spin ordering in MnPSe$_3$ is the inspection of the $M_-$ versus $B$ dependence in the limit of low magnetic fields. To access this limit, the phase-sensitive, resistivity-detected microwave magneto-absorption measurements\cite{Cho2023} have been carried out. In this experiment, the resonance absorption signals have been detected when exposing a MnPSe$_3$ specimen to GHz radiation (selected frequencies in the range of 10~GHz to 64~GHz) upon sweeping the magnetic field applied in the arbitrarily chosen direction but remaining along the layers’ plane. As can be seen in Fig.~\ref{fig:Fig3}a and \ref{fig:Fig3}b, the relatively sharp resonances are observed in the high field range where they disperse almost linearly with $B$. They are recognized as M$_-$ magnon modes, indeed merging those observed in Raman scattering experiments at still higher magnetic fields (see Fig.~\ref{fig:Fig2}b). Importantly, however, the measured GHz resonances significantly broaden, indicating the flattening of the E$_{M-}$ versus B dependence when lowering the magnetic field, and no resonance is observed below 20~GHz. This points towards a small but clearly non-zero ($\approx 0.7$ cm$^{-1}$) onset for the M$_-$ magnon gap at $B=0$.

\section*{Discussion}

The results presented above are fully consistent with the identification of the observed M$_+$ and M$_-$ modes as two low energy magnon gap excitations in MnPSe$_3$ antiferromagnet. The temperature dependence of the energy of the M$_+$ mode presented in Fig.~\ref{fig:Fig1}b is typical for a large class of antiferromagnets \cite{Eibschutz1966}. The solid line in the inset of Fig.~\ref{fig:Fig1}b represents the expected temperature softening of the magnon gap given by a conventional formula: 
$w_{M+}(T) = w_{M+}(1-T/T_N)^{1/3}$, in which we have set $w_{M+} = 14$ cm$^{-1}$ for the magnon gap in the low temperature limit and $T_N = 74$~K for the Neél temparature  in MnPSe$_3$ \cite{wiedenmann1981}. 

Further evidence for the identification of M$_+$ and M$_-$ modes comes out from their characteristic dependencies upon the application of the magnetic field. These dependencies are here discussed in reference to a classical theory of the antiferromagnetic resonance in a biaxial antiferromagnet \cite{Nagamiya1955}. First, we focus on the out-of-plane configuration that is the magnetic field applied perpendicularly to the layers' planes ($B \parallel c$). In this case, all Mn$^{2+}$ spins, regardless of their orientation in the plane, are aligned perpendicularly to $B$. The field-dependent energies of M$_+$ and M$_-$ modes are given by the following expressions \cite{Nagamiya1955}:
 \begin{equation} \label{eq:1}
w_{M-}(B)= w_{M-}
\qquad w_{M+}(B)= \sqrt{(w_{M+})^2 +(g\mu_B B)^2} 
\end{equation}
where $w_{M-}$ and $w_{M+}$ denote the M$_-$ and M$_+$ magnon energies at $B=0~T$ and $g$ is the effective $g$-factor. We thus expect that the $w_{M-}$ remains unaffected by the out-of-plane magnetic field, while $w_{M+}$ increases with $B$, initially quadratically but linearly at high fields. The $w_{M+}(B)$ dependence given in the above formulas is in accordance with the results presented in Fig.~\ref{fig:Fig2}a: The dashed line in Fig.~\ref{fig:Fig2}a has been drawn when setting $w_{M+}=14$~cm$^{-1}$ and $g=2$ for the effective g-factor. We also note that when replacing $B$ by $B\cos({20^{\circ})}$, the above formula describes well the Raman scattering data presented in the Fig.~S2 of the SM\cite{SuppInfo}.

The analysis of the data obtained in the in-plane magnetic field configuration is more complex. As illustrated in Fig.~S4 of the SM, the apparent shapes of $w_{M-}(B)$ and $w_{M+}(B)$ dependencies are different for different orientations of the magnetic field with respect to the crystal axis along which the spins are initially (at $B=0~T$) aligned \cite{Nagamiya1955}. Unfortunately, this axis is not identified in our work, and moreover, the samples used in our experiments are thick and likely composed of grains with different orientations of the in-plane crystal axis. The Raman scattering data shown in Fig.~\ref{fig:Fig2}b correspond to the high field range, above the spin-flop field B$_{sf}$. At sufficiently high fields (see section S4 of the SM\cite{SuppInfo}), independent of the actual orientation of the in-plane magnetic field,  the  $w_{M+}(B)$ and $w_{M-}(B)$ energies, satisfy the following conditions:
\newline

$\sqrt{(g\mu_{B}B)^2 - (w_{M-})^2} \leq w_{M-}(B) \leq \sqrt{(g\mu_{B}B)^2 + (w_{M-})^2}$
\newline

$\sqrt{(w_{M+})^2 - (w_{M-})^2} \leq w_{M+}(B) \leq w_{M+} $
\newline

Expecting a rather small value for the spin-flop field ($B_{sf}=0.7$ T)~\cite{Basnet2022} and thus a small value for the $w_{M-}=0.7$~cm$^{-1}$, one finds that in the high field range $w_{M+}$ is not affected by the applied field ($w_{M+}(B)=w_{M+}$) whereas, $w_{M-}(B)$ depends linearly on the applied field ($w_{M-}(B)=g\mu_{B}B$). This accounts for the observation, as shown in the inset to Fig.~\ref{fig:Fig2}b, of linear with B-field dependence of $w_{M-}(B)$.

Less evident is the interpretation of the data presented in Fig.~\ref{fig:Fig3}a, which results from our approach to trace the $w_{M-}(B)$ dependence in the range of low in-plane magnetic fields, on either side of the spin-flop field. As previously stated, access to this range can be achieved through an alternative experimental approach, resembling the multi-frequency magnetic resonance technique. This method necessitates the utilization of relatively thick specimens, unavoidably composed of grains with different orientations of the in-plane crystal axis. Strictly speaking, the data presented in Fig.~\ref{fig:Fig3}a can not be accounted for by a single $w_{M-}(B)$ trace. The $w_{M-}(B)$  dependence is different for each grain, with different orientations of the crystal axis with respect to the magnetic field (see Fig.~S4b of the SM). Nevertheless, the results shown in Fig.~\ref{fig:Fig3}a can be effectively reproduced when "averaging" the $w_{M-}(B)$ dependencies expected for differently oriented flakes.  
The False color map shown in Fig.~\ref{fig:Fig3}b has been constructed by summing up the multiple $w_{M-}(B)$ dependencies obtained by varying, every $1^\circ$, the respective angle between the applied magnetic field and the axis of the initial spin alignment (certain crystal axis). Each of these traces has been calculated assuming $w_{M-}=0.7$~cm$^{-1}$ for the energy of the M$_-$ mode at zero magnetic field and consistently with the previous analysis, $g=2$ for the effective $g$-factor. Most of these traces concentrate in the upper energy limit given by 
$w_{M+}(B)= \sqrt{(w_{M+})^2 +(g\mu_B B)^2}$ 
which, as can be seen in Fig.\ref{fig:Fig3}b, reproduces the experimental data fairly well.  As expected the resonances are broad in the limit of low magnetic fields but relatively narrow at higher fields, where they disperse linearly with the magnetic field and match those traced with magneto-Raman scattering (see Fig.~\ref{fig:Fig2}b).

Having established two low energy magnon gaps in MnPSe$_3$, we identify this antiferromagnet as a biaxial system with two anisotropy fields ($B_{D\perp}$ and $B_{D\parallel}$) or energies ($D_{\perp}$ and $D_{\parallel}$): $D_{\perp}=g\mu_B B_{D\perp}$ and $D_{\parallel}=g\mu_B B_{D\parallel}$. In agreement with previous reports\cite{Basnet2022}, the out-of-plane anisotropy $D_{\perp}= 0.3$~cm$^{-1}$ restricts the in-plane spin alignment and accounts for the upper magnon gap: $w_{M+}=\sqrt{2E_{exch}D_{\perp}}=14$~cm$^{-1}$, where $E_{exch}$ is the mean exchange energy. On the other hand, the reported here, non-zero gap $w_{M-}=0.7$~cm$^{-1}$ for the lower magnon branch indicates the presence of the additional in-plane anisotropy $D_{\parallel}$ ($w_{M-}=\sqrt{2E_{exch}D_{\parallel}}$). Although rather weak: $D_{\parallel}=D_{\perp} \times (w_{M-}/w_{M+})^2= 7.5 \times 10^{-4}$~cm$^{-1}$,  $D_{\parallel}$ imposes spins to be aligned along certain in-plane crystal axis. This axis remains unidentified in the present work which calls for further, experimental and theoretical studies of MnPSe$_3$ antiferromagnet.




 



\section*{Concluding remarks}

We have employed Raman scattering, far-infrared transmission, and microwave absorption measurements, carried as a function of temperature and of the magnetic field, to identify magnon gap excitations in MnPSe$_3$ van der Waals antiferromagnet. The observation of two distinct magnon gap excitations supports the identification of MnPSe$_3$ as a biaxial antiferromagnet with non-negligible in-plane anisotropy. Relevant parameters (fundamental magnon gap energies, $g$-factor, spin-flop field, anisotropy energies) are also estimated.

\section*{Methods}

We utilized commercially available thick and large area ($\approx25$~mm$^2$ $\times$ 500~$\mu$m) MnPSe$_3$ crystals for FIR-transmission and Raman scattering measurements and relatively thin and small size crystals for GHz absorption measurements. Magneto-optical experiments were performed in both the Faraday and Voigt configurations with the magnetic field (from a superconducting solenoid magnet) applied perpendicular and parallel to the layers' plane respectively. In the case of Raman scattering measurements, the sample was excited by a semiconductor-based laser ($\lambda$ = 515 nm) at a tilt of 20$^\circ$ with respect to the surface normal\cite{DipankararXiv}. The laser beam was focused on the sample using a microscope objective with a numerical aperture of 0.83. The scattered signal was collected using the same objective and was analyzed by a 0.7m grating spectrometer equipped with a liquid nitrogen-cooled charge-coupled device (CCD). A set of reflection-based Bragg filters was employed in both the excitation and collection paths to effectively eliminate any unwanted laser line and to reject backscattered laser. For far infrared magneto-transmission measurements\cite{DipankararXiv}, the radiation emitted from a mercury lamp was modulated using a Bruker Vertex 80v Fourier-transform spectrometer. The modulated radiation was delivered to the sample through light-pipe optics and subsequently detected by a composite bolometer placed directly behind the sample. Magneto-absorption experiments in the GHz frequency range were performed utilizing the modulation technique\cite{Cho2023}. The radiation emitted by an Agilent gigahertz source was transmitted to the sample through a coaxial cable. The sample was positioned in the proximity of a resistive thermometer, which served as a bolometer. When the sample effectively absorbs the incident radiation, it results in a reduction of the radiation reaching the resistor. This causes a change in the resistor's temperature, which can be detected by monitoring the variation in its resistance. The alteration in resistivity is detected through changes in modulation response amplitude, while the delay in the resistivity response, owing to the finite thermalization time, is observed through the modulation response phase. The change in the modulation response phase with respect to the off-resonance phase is plotted in Fig.~\ref{fig:Fig3}a.

\section*{Data Availability}
The datasets generated during and/or analysed during the current study are available from the corresponding author on reasonable request.


\section*{Acknowledgements}

The work has been supported by the EC Graphene Flagship Core 3 project. M.P and M.B acknowledge the support from the Foundation for Polish Science (MAB/2018/9 Grant within the IRA Program financed by EU within SG OP Program).

\section*{Author contributions statement}

D.J., J.-Ph., B.A.P., M.O., C.F., and M.P. conceived the project and experiments. D.J., D.V., I.M., P.K., C.W.C., I.B., and M.B. conducted the experiments and analyzed the results. The manuscript, initially written by D.J. and M.P., has been reviewed by all authors.

\section*{Additional information}

\textbf{Accession codes}

\textbf{Competing interests:} 
The author(s) declare no competing interests.


\newpage
\pagenumbering{gobble}

\begin{figure}[htp]
\includegraphics[page=1,trim = 18mm 18mm 18mm 18mm,
width=1.0\textwidth,height=1.0\textheight]{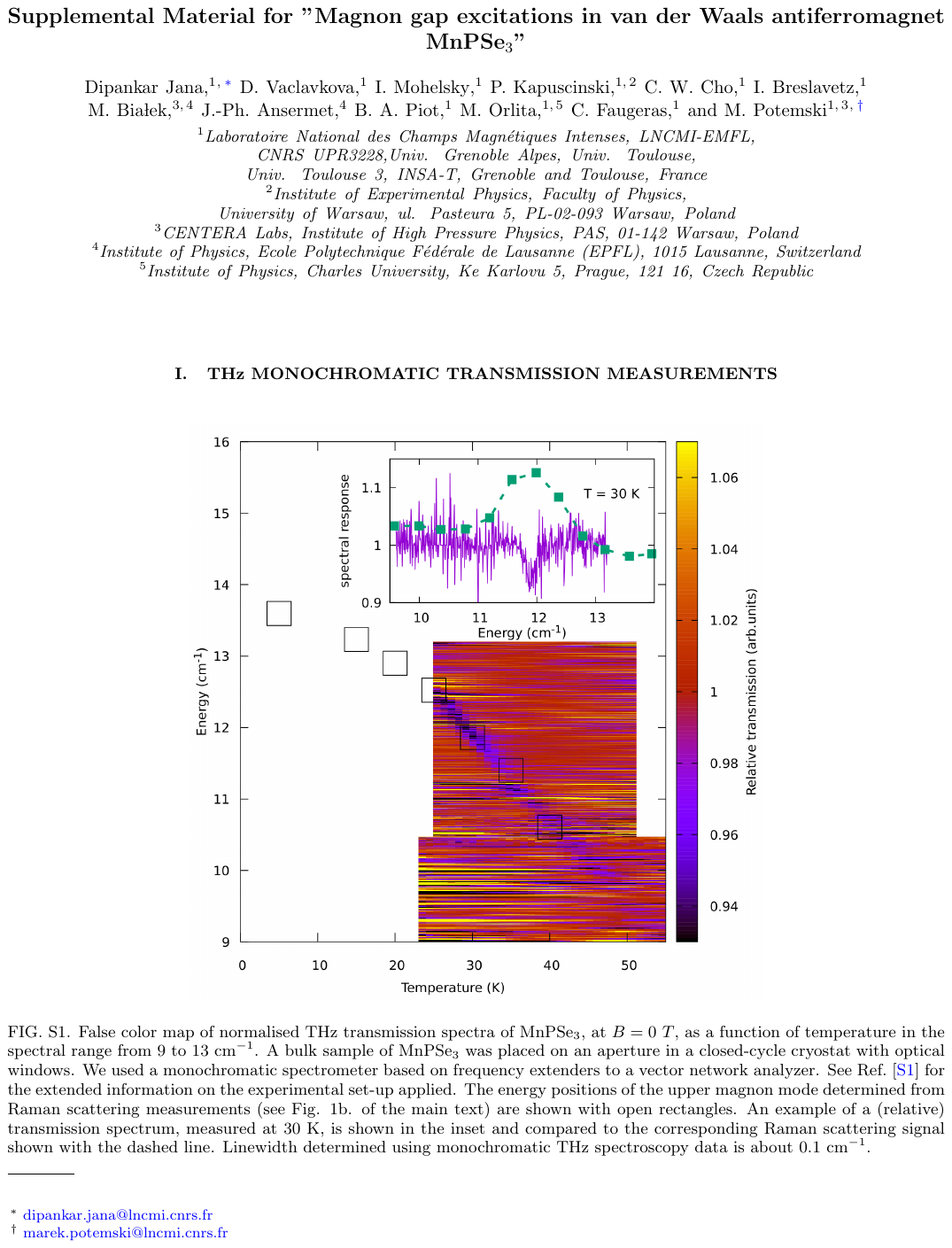}
\end{figure}

\newpage

\begin{figure}[htp]
   \includegraphics[page=2,trim = 18mm 18mm 18mm 18mm,
width=1.0\textwidth,height=1.0\textheight]{MnPSe3_Magnon_SM.pdf}
\end{figure}
\newpage

\begin{figure}[htp]
   \includegraphics[page=3,trim = 18mm 18mm 18mm 18mm,
width=1.0\textwidth,height=1.0\textheight]{MnPSe3_Magnon_SM.pdf}
\end{figure}

\begin{figure}[htp]
   \includegraphics[page=4,trim = 18mm 18mm 18mm 18mm,
width=1.0\textwidth,height=1.0\textheight]{MnPSe3_Magnon_SM.pdf}
\end{figure}

\end{document}